# Additional details about the thermal design error in *'A Photon Thermal Diode'*

Chris Dames[1,*] and Zhen Chen[1]

Affiliations as of the original publication:
[1]Department of Mechanical Engineering, University of California, Berkeley, CA 94720, USA
*email: cdames@berkeley.edu

This note supplements a Reply [1] to a Comment [2] which identified a fundamental symmetry error we made in the original experimental study of a photon thermal diode [3]. For thermal specialists, here we provide some additional details about the thermal estimate originally used to justify omitting the cold-side collimator, and the reasons for its failure.

The schematics in Figure 1 below help explain the issues. For an ideal experimental design with two blackbodies (BBs) and two graphite plate thermal collimators, it is obvious that the "thermal polarity" can be reversed by flipping either the BB temperatures (Fig. 1A), or the thermal diode test section (Fig. 1B). Of critical importance here, for proper symmetry of Fig. 1B the test section to be flipped should include everything between the BBs: the pyramidal mirrors plus any graphite plate(s).

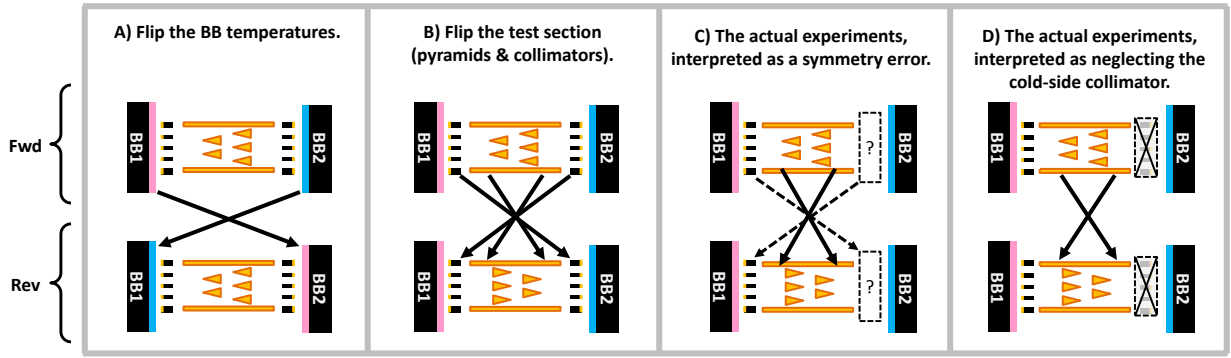

Figure 1. Symmetries of the experiments in question. (A,B) Two equivalent schemes to switch between forward (top row) and reverse bias (bottom row). In the actual experiments (C,D: two different interpretations) only one graphite plate "thermal collimator" was installed, always at the hot side. (C) As pointed out in the Comment [2], this is a fundamental symmetry error. (D) During the original design [3] we considered the missing collimator as an experimental simplification, mistakenly concluding at the time that it would introduce negligible errors. This note explains the estimate we had used to justify (D), and why it is wrong.

The actual experiments with thermal collimation are depicted schematically here in Fig. 1(C,D). By retaining only the hot-side collimator, this configuration is not equivalent to the ideal reversal scheme of Fig. 1(B), and thus introduces a fundamental symmetry error [2]. Although we were aware of the symmetry issue during the initial experimental design, for reasons of experimental practicality we had chosen to forego the cold-side collimator at the same time we decided to use a flat black plate rather than a graphite cavity for BB2 [4]. We judged this acceptable based on a flawed thermal estimate, which we believed showed that omitting the cold-side collimator (Fig. 1D) would cause only minor errors in the overall heat transfer.

The key question is how much error in the forward and reverse heat flows is expected from omitting the cold-side collimator (Fig. 1D vs. Fig. 1B). Underlying the argument is the strong quartic scaling of the blackbody emissive power, $E_b = \sigma T^4$. Since during these experiments $T_{BB1} \approx 2T_{BB2}$, the heat injected into the test section from BB2 is much less than that from BB1, by a factor of approximately 16. More precisely, using the actual aperture area (40.3 cm$^2$) and typical temperatures ($T_H$=561 K, $T_C$=283 K) of the experiments, the respective powers under a black approximation are $Q_{injected,1}$ = 22.65 W and $Q_{injected,2}$ = 1.47 W, a factor of 15.4. Thus, from this gross power perspective, the cold side BB is only around 6.5% as important as the hot side BB, an observation which had suggested to us that the impact of the cold-side collimator was similarly less than that of the hot-side collimator.

| Configuration | Pyramids | Collimators | $Q_{net,Fwd}$ [W] | $Q_{net,Rev}$ [W] | Difference (Fwd.-Rev.) [W] | Average [W] | Difference / Average |
|---|---|---|---|---|---|---|---|
| Controls | Yes | None | 10.02 | 10.05 | 0.03 | 10.04 | 0.3% |
| Collimator 1 | Yes | Hot-side only | 9.50 | 8.57 | 0.93 | 9.04 | 10.3% |
| Difference (Controls - Col.1) [W]: | | | 0.52 | 1.48 | | | |

Table I. Summary of measured heat flows, corresponding to Fig. 3C of [3]. For comparison, the detection limit is $\delta Q_{min} = 0.08\,\text{W}$, and the heat flows injected from the two BBs are $Q_{injected,1}$ = 22.65 W and $Q_{injected,2}$ = 1.47 W.

To make the discussion more quantitative we refer to measurements from Fig. 3C of [3], summarized here in Table I. See also Figure 2 below. Adding the hot-side collimator had only a modest impact on the net heat flow as compared to the no-collimator control experiments, with reductions of $\delta Q_{coll.,Hot} \approx 0.52$ W – 1.48 W, which is only 2.3% - 6.5% of $Q_{injected,1}$. Given this measured reduction in heat flow caused by inserting the *hot*-side collimator next to *BB1* (Figure 2(a→b) here), as a first approximation it had seemed reasonable to estimate the additional reduction expected from inserting the *cold*-side collimator next to *BB2* by applying a similar 2.3% - 6.5% scaling (Figure 2(b→c) here), namely,

$$\delta Q_{coll.,Cold} \approx (2.3\% - 6.5\%) \cdot Q_{injected,2} \approx 0.03 \text{ - } 0.10 \text{ W}.$$

That is, if the cold-side collimator had been included, this line of thinking suggests that the net heat flow would have been reduced by an additional ~0.03 - 0.10 W. Since this is small compared to the measured rectification $\delta Q_{rect.} = 0.93$ W, this estimate suggests that omitting the cold-side collimator in the actual experimental configuration of Fig. 1(D) should not dramatically change the rectification effect as compared to the proper mirror symmetry of Fig. 1(B).

However, the above argument neglected a crucial effect. Consider again the Landauer-Büttiker flow diagrams of Fig. 2 here. The above argument used the measured $\delta Q_{coll.,Hot}$, interpreted as the effect of the hot-side collimator in transmitting and reflecting $Q_{injected,1}$, as a rational basis to make an analogous estimate for $\delta Q_{coll.,Cold}$, the effect expected of the cold-side collimator in transmitting and reflecting $Q_{injected,2}$. But these phenomena are not localized in this way. After some fraction of $Q_{injected,1}$ passes through the hot-side collimator and the pyramidal mirrors, it still must transmit through the cold-side collimator before terminating in BB2. At the plane where the cold-side collimator would be, we estimate that what remains of $Q_{injected,1}$ is ~10 W. In this case, applying a 2.3% - 6.5% reduction as described above now corresponds to $\delta Q_{coll.,Cold,onQinjected1} \approx 0.23 - 0.65$ W. This estimate is still quite crude because it neglects all the additional scatterings back and forth among the three major internal components (pyramid test section and both collimators), though in principle these could be accounted for using infinite sums or a transmission matrix method. The essential point here is that this range of estimated $\delta Q_{coll.,Cold,onQinjected1}$ is not negligible compared to the main rectification result of $\delta Q_{rect.} = 0.93$ W, so we now conclude that omitting the cold-side collimator likely had a significant effect [1,2], invalidating several key measurements of the original work [3].

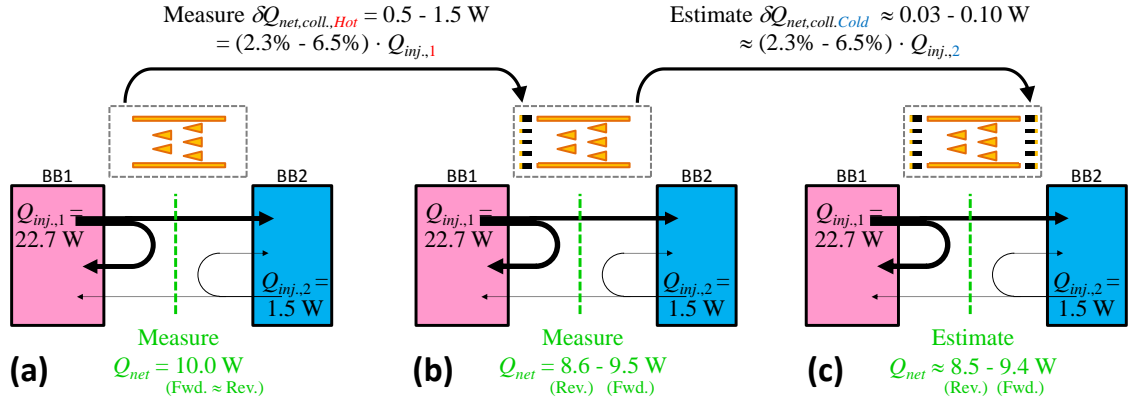

Figure 2. Measured and estimated impacts of the thermal collimators on the overall heat currents. See also Table I. (a→b) Measurements showed that adding the thermal collimator near BB1 reduced the total heat flow by 0.5 - 1.5 W, corresponding to 2.3% - 6.5% of $Q_{injected,1}$. (b→c) To estimate the additional impact on $Q_{net}$ expected from a cold-side collimator near BB2, we had applied this same relative correction to $Q_{injected,2}$. Since $Q_{injected,2} << Q_{injected,1}$, the estimated correction is less than ~0.1 W, negligible compared to the measured rectification of 0.9 W. The fatal problem with this argument is that it fails to consider the additional impact of the cold-side collimator in backscattering the remaining portion of $Q_{injected,1}$ after it traverses the hot-side collimator and the pyramids.

---